\documentclass[12pt,preprint]{aastex}

\shorttitle{Plasma Anomaly in Cygnus}
\shortauthors{Whiting et al}

\begin{document}

\title{Confirmation of a Faraday Rotation Measure Anomaly in Cygnus}
\author{Catherine A. Whiting, Steven R. Spangler, and Laura D. Ingleby$^1$\altaffiltext{1}{Present address, Department of Astronomy, University of Michigan}}
\affil{Department of Physics and Astronomy, University of Iowa, Iowa City, IA 52242}
\author{L. Matthew Haffner}
\affil{Department of Astronomy, University of Wisconsin, 475 North Charter St., Madison, WI 53706}
\begin{abstract}
We confirm the reality of a reversal of the sign of the Faraday Rotation Measure in the Galactic plane in Cygnus \citep{Lazio90}, possibly associated with the Cygnus OB1 association. The rotation measure changes by several hundred rad/m$^2$ over an angular scale of $2-5^{\circ}$. We show that a simple model of an expanding plasma shell with an enhanced density and magnetic field, consistent with observations of H$\alpha$ emission in this part of sky, and physically associated with a superbubble of the Cygnus OB1 association, can account for the magnitude and angular scale of this feature. 
\end{abstract}


\section{Introduction}
Faraday rotation is one of the best diagnostics of the plasma properties of the interstellar medium, in particular the strength and direction of the interstellar magnetic field \citep{Clegg92,Frick01,Haverkorn04,Vallee04,Haverkorn06}. When radio emission from a source of polarized radio waves propagates through a plasma such as the interstellar medium, the measured polarization position angle $\chi$ is given by 
\begin{equation}
\chi = \chi_0 + \left[ \left( \frac{e^3}{2 \pi m_e^2 c^4} \right) \int n \vec{B} \cdot \vec{ds} \right] \lambda^2
\end{equation} 
where $\chi_0$ is the intrinsic polarization position angle, assumed independent of wavelength ($\lambda$), $n$ is the plasma density, $\vec{B}$ is the vector magnetic field, and $\vec{ds}$ is an incremental pathlength interval along the line of sight.  The quantities $e$, $m_e$, and $c$ are the usual fundamental physical constants of the elementary electrical charge, the mass of an electron, and the speed of light.  The quantity within square brackets is termed the {\em rotation measure} ($RM$) and is the quantity which constitutes a plasma diagnostic of the interstellar medium (ISM).  Equation (1) is in the cgs system of units.  Conversion of the rotation measure to the more common SI units of radians/m$^2$ is done by multiplying the cgs value by a factor of $10^4$.  

The rotation measure is a signed quantity, depending on the density-weighted line of sight component of the magnetic field.  The sign of the $RM$ is of physical interest because it diagnoses the net polarity of the magnetic field along the line of sight.  As may be seen in equation (1), an increase in $\chi$ with increasing $\lambda$ indicates that $RM$, and thus the path-averaged magnetic field, are positive.  

Measurements of rotation measure have been made along hundreds of lines of sight through the galaxy, using many radio telescopes.  These observations have been used to determine the large scale structure of the Galactic magnetic field \citep[e.g.][]{Clegg92,Vallee04} as well as properties of turbulence in the ISM \citep{Lazio90,Minter96,Haverkorn04}.

This paper is particularly concerned with results reported by \cite{Lazio90}.  Polarization measurements at three frequencies were made with the Very Large Array (VLA)\footnote{The Very Large Array is an instrument of the National Radio Astronomy Observatory.  The NRAO is a facility of the National Science Foundation, operated under cooperative agreement with Associated Universities, Inc.}of 8 extended radio sources observed through the Galactic plane in Cygnus, in the vicinity of the Cygnus OB1 association. Previous radio observations had shown this region to be one of strong radio wave scattering, presumably due to plasma turbulence associated with the Cygnus OB1 association \citep{Spangler88,Fey89,Spangler98}. The distance to the Cygnus OB1 association is estimated to be 1.8 kpc \citep{Nichols93}, and this value for the distance will be employed in this paper.  The main point of the observations of \cite{Lazio90} was to determine properties of the plasma turbulence on large spatial scales, manifest by differences $\Delta RM$ in the rotation measures on adjacent lines of sight to different parts of a background extragalactic radio source, and between lines of sight to different sources. The observations of \cite{Lazio90} were consistent with a Kolmogorov spectrum of irregularities in the plasma around the Cygnus OB1 association, extending from spatial scales as small as $10^8 - 10^9$ cm, which are responsible for diffractive interstellar scattering, to as large as $10^{17} - 10^{18}$ cm, scales which cause Faraday rotation changes between adjacent lines of sight. 

The magnitude of Faraday Rotation in this part of the Galactic plane was found to be sufficiently large to cause difficulty in extracting RM values from the observations at frequencies of 1.44, 1.65, and 4.88 GHz made by \cite{Lazio90}. The basic problem is that for $RM$s of several hundred rad/m$^2$, there can be several turns of the position angle between 1.65 and 4.88 GHz, and even a turn (or more) between 1.44 and 1.65 GHz.  This is illustrated graphically in Figure 3 of \cite{Lazio90}. A technique was employed whereby the optimum set of values of $RM$ consistent, in a least-squares sense, with the position angle measurements was retrieved.  In the cases of four sources, 2004+369, 2005+368, 2007+365, and 2011+360, secure enough values for the average $RM$ to the source were obtained to permit ``unambiguous'' rotation measures to be reported in Table 3 of \cite{Lazio90}.  
The $RM$ values reported were all positive, with values between 228 and 850 radians/m$^2$.   

\cite{Clegg92} presented polarization observations yielding $RM$ for 56 extragalactic radio sources viewed through the Galactic plane, of which 4 were in the vicinity of the Cygnus OB1 association.  The $RM$ values reported were large (several hundred radians/m$^2$), and {\em negative}. There were no sources in common between the samples of \cite{Lazio90} and \cite{Clegg92}. This apparent large change in the $RM$ (from several hundred radians/m$^2$ positive to a similar magnitude, but negative) within a small angular distance suggested either (a) an error in the observations and/or data reduction procedure by \cite{Lazio90}, so that the sign change did not exist, or (b) the existence of a remarkable plasma structure in the Galactic plane in this region, which could cause such a large, anomalous $RM$.  We refer to this large change in the Galactic Faraday rotation measure over a small angular scale as a Rotation Measure Anomaly. A partial argument against possibility (a) was available even at the time of \cite{Clegg92}.  Processing of the polarization position angle measurements made and published by \cite{Lazio90} through the $RM$ algorithm of \cite{Clegg92} yielded the $RM$ values given in Table 3 of \cite{Lazio90}.  

The purpose of this paper is to report observations made to confirm the large, positive $RM$ values of \cite{Lazio90}.  Observations were made with the VLA at two, widely separated frequencies within the 5 GHz RF (radiofrequency) bandpass.  Observations were made at this frequency because $RM$ values of the magnitude reported by \cite{Lazio90} and \cite{Clegg92} would cause measurable rotation of several degrees of position angle or more between the two bands, but there would be no `` n-$\pi$ ambiguity''. As will be seen below (Section 3), these new observations do confirm the values reported in \cite{Lazio90}, indicating that a strongly Faraday-rotating structure is along the line of sight to these sources, and probably associated with the Cygnus OB1 association.  

The outline of this paper is as follows.  In Section 2 we describe the polarimetric observations made with the VLA at 5 GHz (C band), and optical observations made with the Wisconsin H-Alpha Mapper (WHAM) to investigate the plasma along this line of sight.  In Section 3, we compare the $RM$ values obtained here with the values of \cite{Lazio90}, and see that they are consistent.  In Section 4, we develop a simple model for the plasma structure of an expanding plasma shell, presumably a stellar bubble, associated with the winds and UV radiation of the Cygnus OB1 association, and show that this model is capable of reproducing the magnitude and angular scale of the observed $RM$ anomaly.  Finally, Section 5 summarizes and concludes.

\section{Observations}
\subsection{VLA Polarimetric Observations}  Radio polarization measurements of four sources from the sample of \cite{Lazio90}  were made on 13-14 December, 1999.  The observations were made between 19:46 UT on the 13th to 3:42 UT on the 14 of December.  The sources observed are given in Table 1.  Column 1 gives the J2000 name of the source, and column 2 gives the B1950 name used in \cite{Lazio90}.   The remaining columns are described in Section 3 below.  
\begin{deluxetable}{ccrrr}
\tabletypesize{\small}
\tablecaption{Sources Observed and Faraday Rotation Measurements\label{tbl-1}}
\tablewidth{0pt}
\tablehead{\colhead{Source (J2000)} & \colhead{Source (B1950)} & \colhead{Comp.} & \colhead{$\Delta \chi (\circ)$} & \colhead{$RM$ (rad/m$^2$)}}
\startdata
2007+369 & 2005+368 & Np & $20.1 \pm 0.7$ & $689 \pm 24$  \\
\ldots & \ldots & Sf & $21.4 \pm 0.4$ & $732\pm 12$  \\
2009+367 & 2007+365 & Sf & $7.8 \pm 1.1$ & $269 \pm 40$  \\
\ldots& \ldots& Np & $5.9 \pm 1.4$ & $201\pm 48$  \\
2013+361 & 2011+360 & Sp& $4.5 \pm 0.9$ & $154 \pm 30$  \\
\ldots& \ldots& cc & $10.9 \pm 6.2$ & $373 \pm 213$  \\
\ldots& \ldots&Nf & $ -1.1 \pm 6.6$  & $-37 \pm 226$  \\
2015+364 & 2013+362 & Sp& $-0.2 \pm 0.2$ & $-7.5 \pm 6.2$  \\
\ldots& \ldots& Nf & $-3.3 \pm 0.4$ & $-114 \pm 15$  \\
\enddata
\end{deluxetable}
Three of the sources listed in Table 1, 2007+369, 2009+367, and 2013+361, had rotation measures listed as ``unambiguous'' by \cite{Lazio90}.  The fourth source, 2015+364, was not so listed but was observed in the present project because of the high polarized intensity, which facilitated a comparison of our observations and results with those of \cite{Lazio90}.  

Observations were made at two frequencies, 4585 and 4885 MHz, within the C band RF bandpass.  At both frequencies a 50 MHz  intermediate frequency (IF) bandwidth was received and correlated.  The 300 MHz separation between the two bands is about the maximum possible within the overall C band front-end bandpass, while still having similar gain and acceptable system temperature. For the purposes of this project, we consider polarization measurements at two frequencies as adequate for measuring Faraday rotation, rather than three frequencies as conventionally prescribed. As discussed in the preceding section, given the rotation measure values published by \cite{Lazio90}, the Faraday rotation between the closely-spaced frequencies of 4585 and 4885 MHz should be a small fraction of $180^{\circ}$.  In this case, there is no possibility of the polarization position angle rotating through $180^{\circ}$ or more between the two frequencies of observation (the ``$n \pi$ ambiguity'').    
At the time of the observations, the VLA was in the B array.  

In addition to the four program sources given in Table 1, we also observed the source 2007+404 as a phase and instrumental polarization calibrator, and 3C138 for purposes of calibrating the polarization position angles.  The procedures of polarization calibration and mapping and analysis of the data are essentially identical to those in previous VLA polarization projects of ours, such as \cite{Lazio90}.  

As in those previous investigations, the basic data products with which we worked were maps of the Stokes parameters Q and U, and maps of the polarization position angle $\chi$ and the polarized intensity $L$, 
\begin{eqnarray}
\chi = \frac{1}{2} \tan^{-1} \left( \frac{U}{Q} \right)  \\ \nonumber
L = \sqrt  (Q^2 + U^2 )
\end{eqnarray}
Since the maps were  made from 5 GHz data taken in the B array,  the angular resolution of the maps was about 1.9 arcseconds (FWHM of synthesized beam).  

The procedure in measuring Faraday rotation from these observations was as follows.  Maps of $\chi$ and $L$ were made at both frequencies.  The maps of $\chi$ at the two frequencies were subtracted from each other, giving a map of 
$\Delta \chi$, where the definition and significance of $\Delta \chi$ are 
\begin{eqnarray}
\Delta \chi = \chi(4585) - \chi(4885) = RM \left[ \lambda_{4585}^2 - \lambda_{4885}^2 \right]  \\  \nonumber
RM=\frac{\Delta \chi}{\lambda_{4585}^2 -  \lambda_{4885}^2}  \\ \nonumber
  = 34.3 \Delta \chi (\circ)  \mbox{ rad/m}^2
\end{eqnarray}
In equation (3) $\chi(4585)$ is the polarization position angle measured at a frequency of 4585 MHz, $\lambda_{4585}$ is the corresponding wavelength at that frequency, and $\Delta \chi ( \circ )$ is the difference in the polarization position angles at 4585 and 4885 MHz, in degrees. In the first two lines of equation (3), $\chi$ and $\Delta \chi$ are in radians and $\lambda_{4585}$ and $\lambda_{4885}$ are in meters, and $RM$ has its conventional units of rad/m$^2$.  In the convenient formula of the third line of equation (3), $RM$ is also in rad/m$^2$, and $\Delta \chi (\circ)$ is in degrees.  

Illustrations of our data reduction procedures are given in Figures 1 and 2, for the sources 2007+369 and 2015+364, respectively.
\begin{figure}[h]
\vspace{0.4cm}
\epsscale{0.60}
\includegraphics[angle=-90,width=18pc]{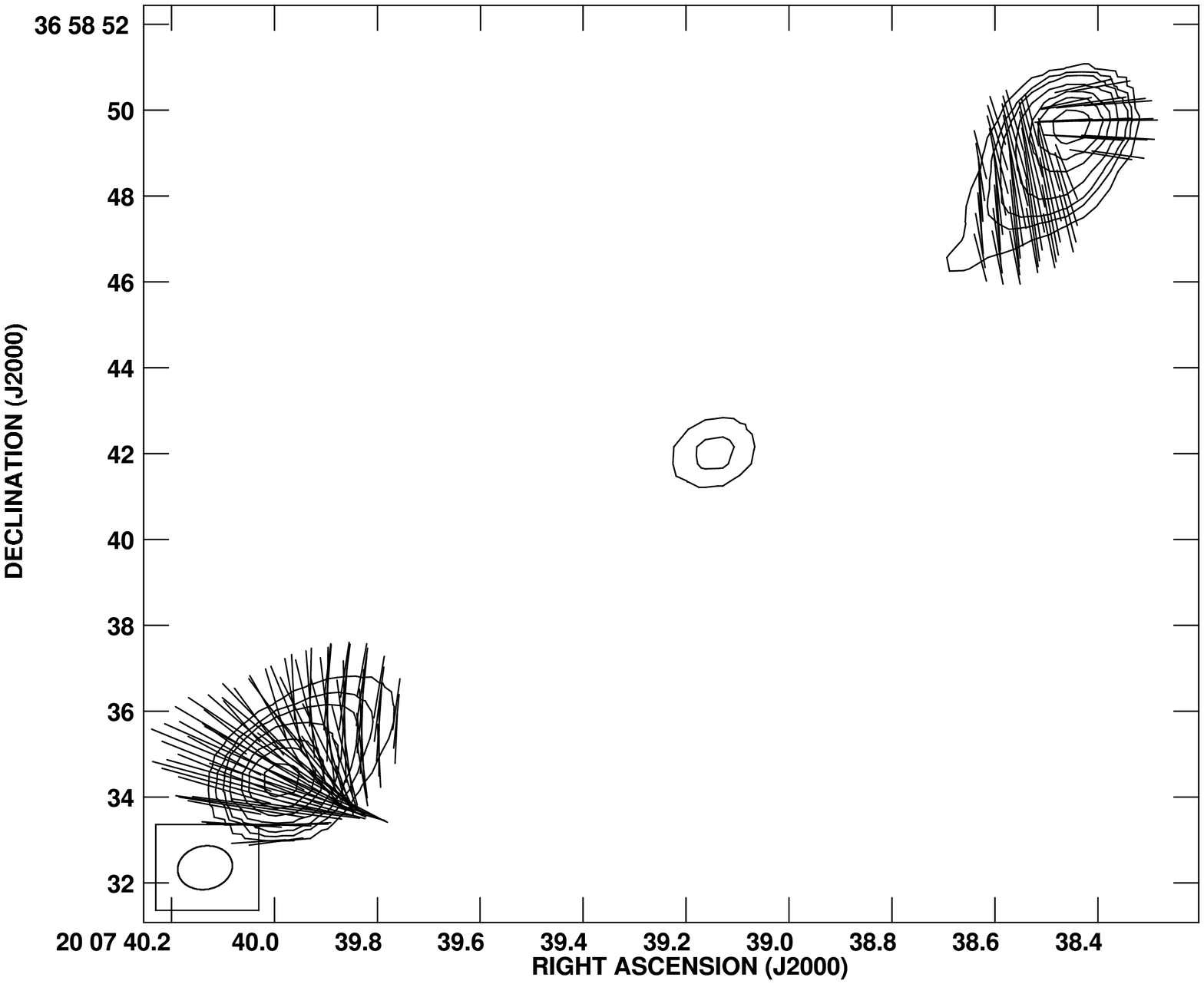}
\includegraphics[angle=-90,width=18pc]{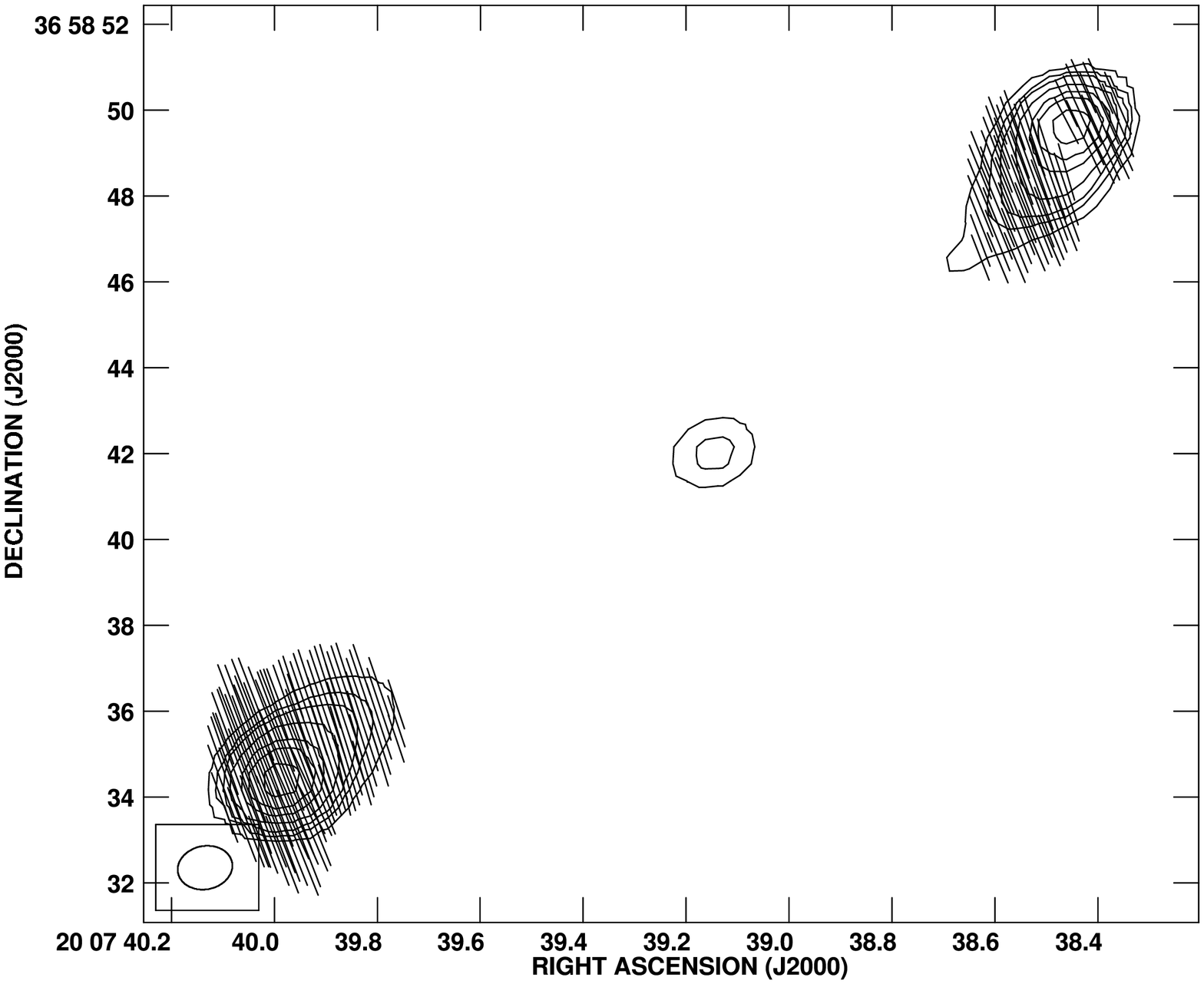}
\caption{Polarization maps of the radio source 2007+369 at a frequency of 5 GHz.  (a; left) Shows the map of the polarization position angle $\chi$ at a frequency of 4.885 GHz.  The contours are those of total intensity, and are given at -3,-1,1,3,5,10,20,30,50,70, and 90 \% of the peak intensity, which is 3.54 mJy/beam.  The restoring beam is shown in the lower left corner, and is $1.9 \times 1.9$ arcseconds.  (b; right) Shows the distribution of position angle difference $\Delta \chi$ between the position angles at 4585 and 4885 MHz}
\end{figure}
\begin{figure}[h]
\vspace{0.4cm}
\epsscale{0.60}
\includegraphics[angle=-90,width=18pc]{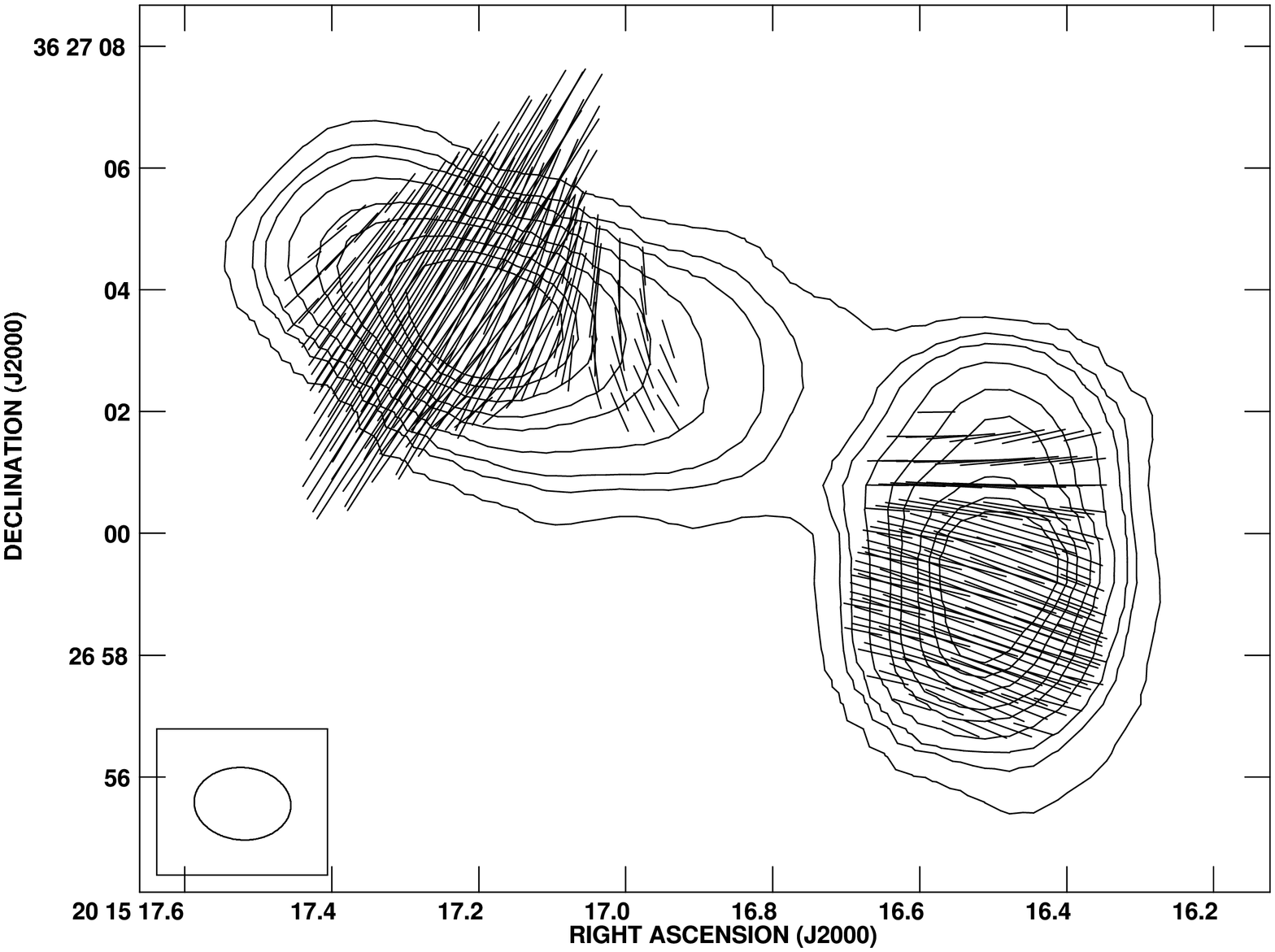}
\includegraphics[angle=-90,width=18pc]{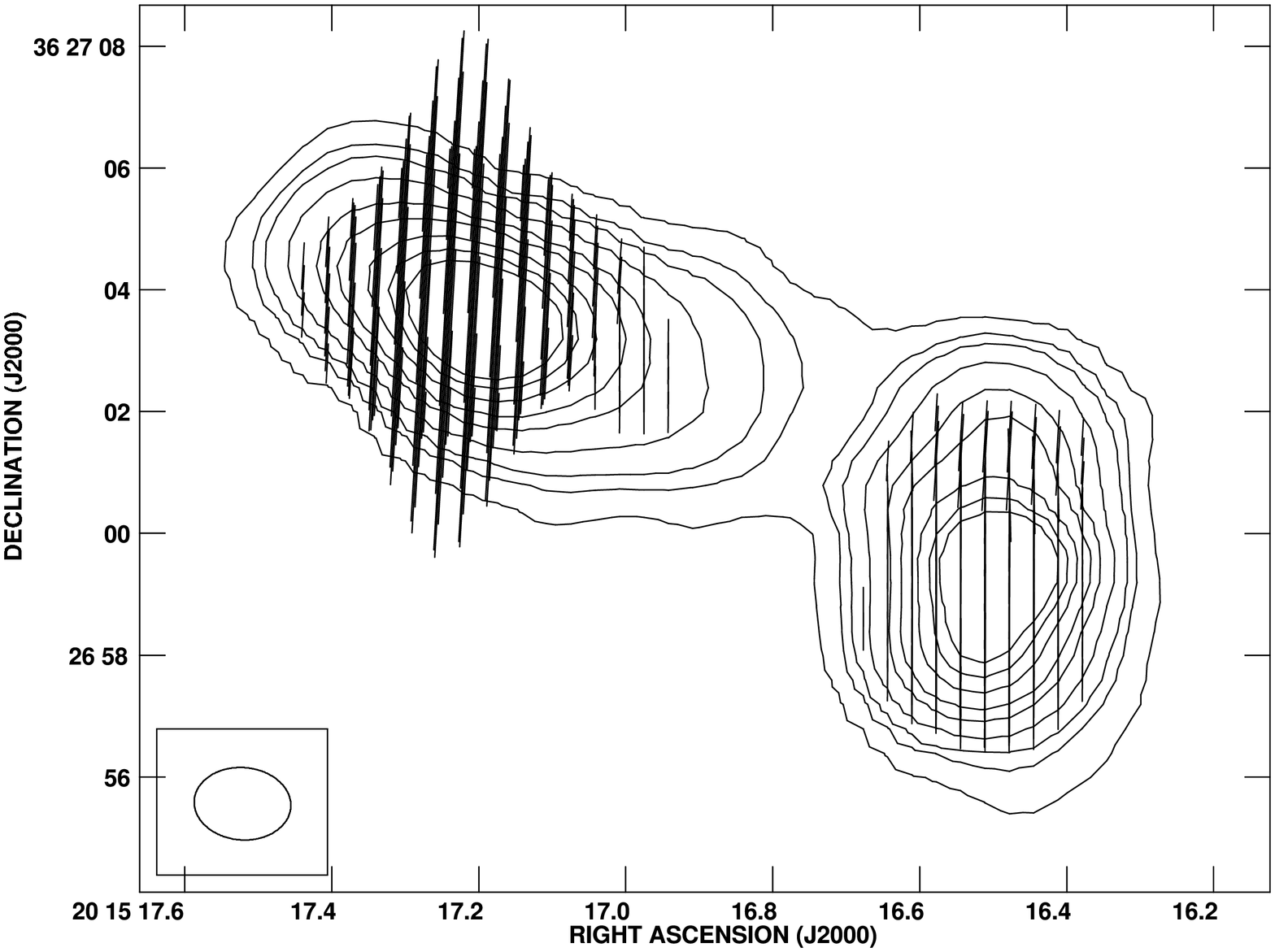}
\caption{Same as Figure 1, except for 2015+364. The peak intensity of the map is 7.11 mJy/beam.}
\end{figure}
In each figure, panel (a) shows the polarization position angle map, with variation reflecting the intrinsic structure of the radio source.  A comparison of these maps with the corresponding ones in Figure  2 of \cite{Lazio90} shows entirely satisfactory agreement, even though the observations of \cite{Lazio90} were made in the C array, and had roughly 1/3 the resolution of the present maps.  This statement is true for all four sources in common between the present project and that of \cite{Lazio90}.

The $\Delta \chi$ maps in panel (b) of both figures show a rotation between the two frequencies, which can be related to $RM$ via equation (3).  For both sources,  $\Delta \chi$ is nearly uniform across the source, being about $20^{\circ}$ for 2007+369, and nearly zero for 2015+364 (see Table 1 for exact values).  The data in these figures clearly indicate that large Faraday rotation variations exist across the Cygnus  OB1 region.  

For each source we chose a number of regions ($\geq 1$) in which the polarized intensity was high enough for a reliable position angle measurement.  In each region, we made a measurement of $\Delta \chi$ at the position where $L$ was locally a maximum.  The errors were calculated on the basis of the conventional standard error in the polarization position angle measurement at a single frequency, $\sigma_{\chi}=\frac{1}{2}\frac{\sigma_Q}{L}$.  

The results of these measurements are shown columns 3, 4, and 5 of Table 1.  Column 3 gives the source subcomponent for the measurement (e.g. Np for  ``north preceding'', Sf for ``south following'', and cc for ``central component'', etc), column 4 gives the $\Delta \chi$ measurement for that component, and column 5 gives the rotation measure to that component, calculated with equation (3).  

\subsection{WHAM H$\alpha$ Observations}
We also analysed data from the University of Wisconsin H-Alpha Mapper \citep[WHAM;][]{Haffner03} for this part of the sky. The WHAM spectrometer samples the emission measure ($EM \equiv \int n_e^2ds$) of plasma along the line-of-sight at one-degree spatial resolution, allowing us to trace structures that could be responsible for the RM anomaly.  We examined both total intensity (H$\alpha$ emission integrated over the whole line profile) as well as 20 km/sec-wide channel maps centered from -60 km/sec to +60 km/sec with respect to the local standard of rest.  

\section{Rotation Measures to Four Sources in Cygnus}
In this section we discuss the significance of the new measurements to the four sources in Cygnus, as well as the implications of the already-published measurements of \cite{Lazio90}.  
\subsection{Comparison with Observations from \cite{Lazio90}}
The $RM$ values from this study (Column 5 of Table 1) are compared with those of \cite{Lazio90} in Figure 3. All data from Table 1 of this paper are displayed, with the exception of the measurement of the central component of 2013+361, since the polarization of this component was not measured by \cite{Lazio90}.   
\begin{figure}[h]
\vspace{0.4cm}
\epsscale{0.60}
\includegraphics[angle=-90,width=22pc]{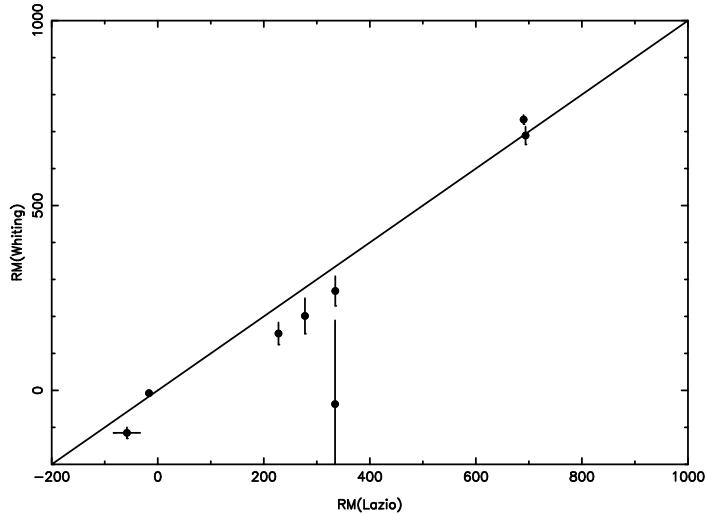}
\caption{Comparison of the $RM$ values emergent from this paper with those of \cite{Lazio90}. }
\end{figure}
The values plotted on the abscissa of Figure 3 are taken from Table 3 of \cite{Lazio90} except for 2015+364.  This was not listed as a source with an ``unambiguous rotation measure'' by \cite{Lazio90}.  Given the small value of $RM$ from our measurements, it was clear that the \cite{Lazio90} measurements were not affected by the ``n $\pi$ ambiguity'', so we fit the position angle measurements from Table 2 of \cite{Lazio90} for the  $RM$ value.  

Figure 3 is the main observational result of this paper.  The measurements from the current investigation are consistent in sign and magnitude with the values published by \cite{Lazio90}.  The apparent discrepancy for the Nf component of 2013+361 is a consequence of the weak polarized intensity of this feature, and the correspondingly large error. Errors for the \cite{Lazio90} measurements are plotted, but for most sources and source components are smaller than the plotted symbols.  

We confirm the large, positive rotation measures for the sources 2007+369, 2009+367, and 2013+361.  Given the accuracy of the  $RM$ values for these sources from \cite{Lazio90}, there is no reason to doubt the sign and magnitude of the fourth source with an unambiguous  $RM$, 2004+369, which also had a large, positive  $RM$ ($\simeq 820$ rad/m$^2$).  

The new measurements also confirm the large spatial gradient in  $RM$ across this region, a change which can be seen to be at least 800 rad/m$^2$ in our data alone, and approximately 1300 rad/m$^2$ when the data sets of this study, \cite{Lazio90}, and \cite{Clegg92} are combined.  

Finally, both our data and the refit position angle data from \cite{Lazio90} show a negative  $RM$ for 2015+364, and thus establish a ``handshake'' between the results of \cite{Lazio90} and \cite{Clegg92}.  We confirm the reality of the rotation measure anomaly identified in \cite{Lazio90} and \cite{Clegg92}, and in Section 4 below we discuss its nature.  

\section{The Source of the RM Anomaly}
In this section, we discuss the nature of the  $RM$ anomaly which has been confirmed by the present observations. We discuss a simple, physically based model for the plasma structure that could produce such a rotation measure anomaly, and also review the evidence for other such objects from observations in the literature. 

To begin the discussion, Figure 4 shows all  $RM$ data in the vicinity of the Cygnus OB1 association.  
\begin{figure}[h]
\vspace{0.4cm}
\epsscale{0.60}
\includegraphics[width=40pc]{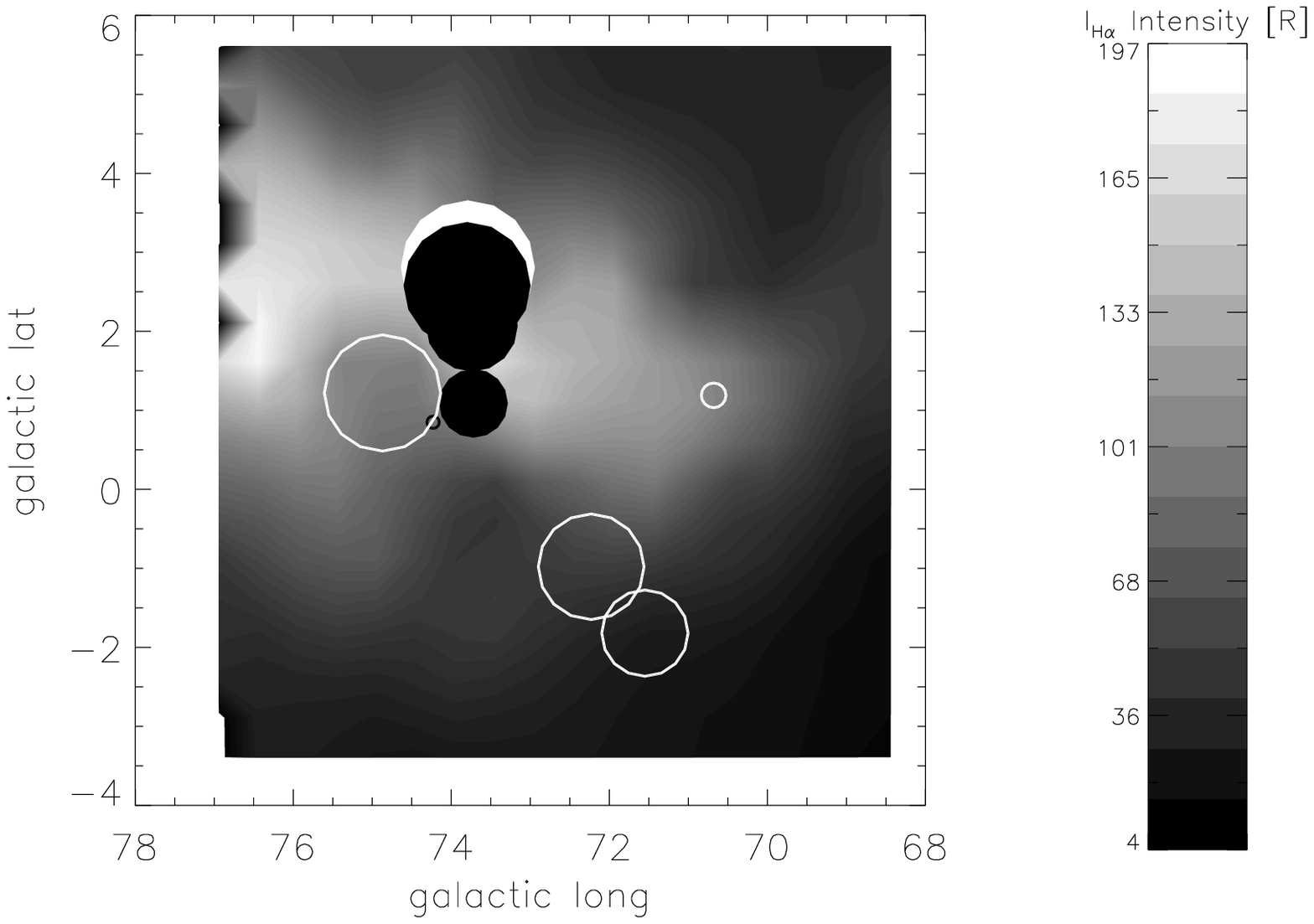}
\caption{ Faraday Rotation in the vicinity of the Cygnus OB1 association.  Positive and negative $RM$ are indicated by solid and open circles, respectively. The diameter is proportional to the logarithm of the absolute magnitude of the $RM$.  The data are from the present paper (4 sources), the source B2004+369 \citep{Lazio90}, and 4 sources from \cite{Clegg92}.  Black symbols portray measurements from this paper, and white symbols are data from \cite{Lazio90} or \cite{Clegg92}. The shaded background illustrates the intensity of the H$\alpha$ emission as mapped by WHAM.  White corresponds to higher intensity, as indicated by the grayscale bar at the right.  The peak intensity is 197 Rayleighs. The WHAM data show an approximate arc of emission with a radius of $2.3^{\circ}$ centered on Galactic coordinates $l=75.0^{\circ}$ and $b=1.0^{\circ}$.  The lower half of the possible arc is absorbed by interstellar extinction in the ``Aquila Rift''. }
\end{figure}
The rotation measure data are represented in the standard format of having solid circular symbols represent positive $RM$, open circles representing negative $RM$, and the plotted diameter of the circle being an indicator of the magnitude of the $RM$. We show our measurements, supplemented by the measurements of \cite{Lazio90} and \cite{Clegg92}. In view of the large range of rotation measures in this part of the sky, the diameter of the plotted symbol in Figure 4 is proportional to the logarithm of the absolute magnitude of $RM$.  Black symbols, both open and filled, represent measurements reported in this paper.  White symbols show data published in \cite{Lazio90} and \cite{Clegg92}.   

The $RM$ measurements are plotted on a background representing the intensity of the integrated H$\alpha$ emission in this part of the sky from the WHAM instrument. The H$\alpha$ emission shows a possible half arc of emission with an angular radius of $2.3^{\circ}$, centered on $l=75.0^{\circ}$ and $b=1.0^{\circ}$. The lower half of this arc is obscured by heavy interstellar extinction in the ``Aquila Rift''. Radio continuum and recombination line observations of this region, summarized in Figure 6 of \cite{Spangler98}, show evidence for ionized gas at and below the Galactic equator at $l = 76^{\circ}$, and extending to $b=1^{\circ}$ at $l = 74^{\circ}$.  Interstellar extinction obscures the emission from this plasma in the WHAM image.   In the discussion in Section 4.1 below, we will assume the physical reality of this shell, and attribute to it some properties of a stellar bubble.  Although its existence and nature are by no means certain, this shell may correspond to the dense shell of shocked and photoionized interstellar gas associated with the bubble (or superbubble) produced by the Cygnus OB1 association \citep{Weaver77}.  A projection of the positions of the O and B stars in the Cygnus OB1 association \citep[taken from][]{Humphreys78} on the H$\alpha$ image shows that most of the stars fall within, or slightly beyond, the possible shell, which mildly corroborates our interpretation of it as the swept-up plasma of an interstellar superbubble. This point was also made in \cite{Nichols93}. 

\subsection{Physical Model of the RM Anomaly}
In this section, we explore a simple physical model for the superbubble shell associated with Cygnus OB1, and see if it can produce a $RM$ anomaly of the sort observed and displayed in Figure 4.  We make no claims as to the uniqueness of this model.  Rather, the goal is to show that structures of the sort known to exist in the vicinity of OB associations {\em can} produce a Faraday Rotation anomaly with a magnitude and angular extent of the sort we observe.  

Figure 5 shows a cartoon illustrating the bubble, which incorporates features from the model of \cite{Weaver77}.  The radius $R_0$ defines the location of the outer shock.  Inside $R_0$ is a shell of thickness $\Delta$, which contains dense, photoionized ISM material.  It is this dense plasma which, in our model, produces the H$\alpha$ shell seen in the WHAM image of Figure 4, and contributes to the enhanced rotation measure.  
\begin{figure}
\vspace{0.4cm}
\epsscale{0.60}
\includegraphics[width=25pc]{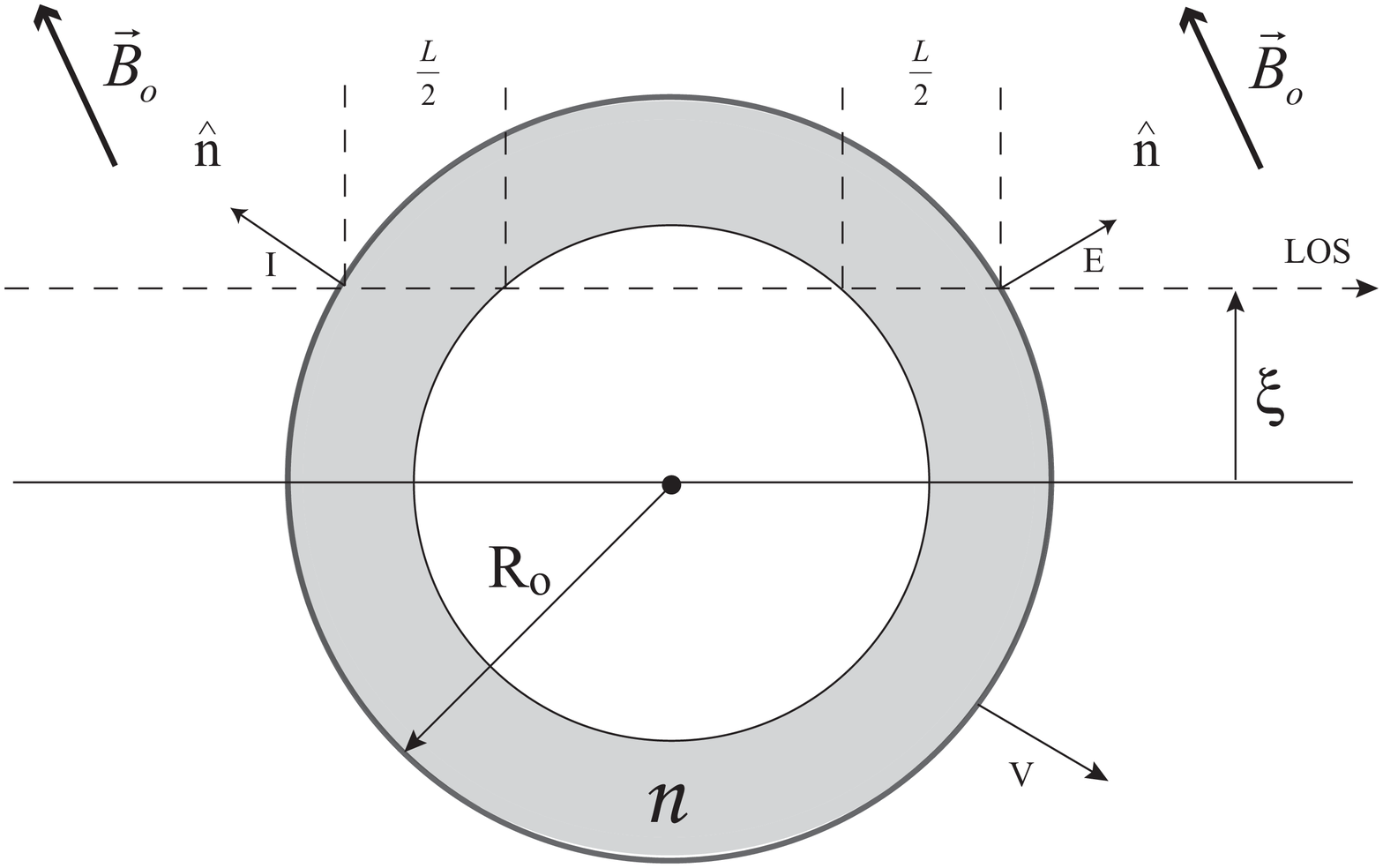}
\caption{Cartoon of a simplified model of a plasma shell associated with a stellar bubble produced by the Cygnus OB1 association.  The cartoon serves to define several of the variables which are used in the discussion in Section 4.1. }
\end{figure} 

The interstellar magnetic field ($\vec{B}_0$) will also be modified by this shell, and it contributes to the enhanced $RM$.  We assume that it is uniform across the region exterior to the bubble, and is modified in the standard way by a magnetohydrodynamic (MHD) shock, in which the component normal to the shock front $B_n$ is unchanged, while the component in the shock plane $B_{\perp}$ is amplified by a factor $X$, $B_{\perp 2} = X B_{\perp 1}$, where 1 and 2 refer to the upstream and downstream regions of the shock.  In the case of a strong shock, $X=4$.  

As indicated in Figure 5, the line of sight (LOS) perforates the shell at two points, I (ingress) and E (egress).  The normal vectors to the shock at these two points are 
\begin{eqnarray}
\hat{n}_I = \frac{\vec{r}_I}{R_0}   \\  \nonumber
 \hat{n}_E= \frac{\vec{r}_E}{R_0}
\end{eqnarray}
where $\vec{r}_I$ and $\vec{r}_E$ are the vectors from the center of the shell to the ingress and egress points.  

The normal and perpendicular magnetic fields in the upstream regions at the ingress and egress points are 
\begin{eqnarray}
\vec{B}_{nI} = (\hat{n}_I \cdot \vec{B}_0)\hat{n}_I    \\  \nonumber
 \vec{B}_{nE} = (\hat{n}_E \cdot \vec{B}_0)\hat{n}_E
\end{eqnarray}
and 
\begin{eqnarray}
\vec{B}_{\perp I} = \vec{B}_0 - \vec{B}_{nI}  \\   \nonumber
 \vec{B}_{\perp E} = \vec{B}_0 - \vec{B}_{nE}
\end{eqnarray}
The net magnetic fields in the shell (the downstream region) at the ingress and egress points are then  given by 
\begin{eqnarray}
\vec{B}_{I2} = \vec{B}_{nI}  + X \vec{B}_{\perp I}  \\   \nonumber
\vec{B}_{E2} = \vec{B}_{nE}  + X \vec{B}_{\perp E}  
\end{eqnarray}
Finally, the $z$ components of the magnetic fields in the shell at the ingress and egress points, 
$B_{Iz}$  and $B_{Ez}$, which determine the Faraday rotation, are obtained by taking the dot product of equations (7) with the unit vector in the $z$ direction (the direction from the source to the observer), $\hat{e}_z$.

As seen from equation (1), the rotation measure is an integral along the line of sight.  In a realistic model of a shell, the line of sight component of the magnetic field $B_z$ as well as the density would vary along the line of sight through the shell.  Given the simplified nature of the model presented here, we approximate the $RM$ as 
\begin{equation}
RM(\xi) = \frac{C n L(\xi)}{2} \left[  B_{zI} + B_{zE} \right]
\end{equation}
where $C$ is the collection of atomic constants within the curved brackets in equation (1),  which has the value $2.631 \times 10^{-17}$ in cgs units, or 0.81 if $L$ is in parsecs and $B_z$ is in microGauss.  The variable $n$ is the assumed uniform plasma density in the shell, $B_{zI}$ and $B_{zE}$ are the downstream (postshock), line of sight components of the magnetic field at the ingress and egress points, respectively.  The length of the chord through the shell is denoted by $L(\xi)$, where $\xi$ is the distance of the line of sight from the center of the shell, as illustrated in Figure 5.  
 Equation (8) assumes that all plasma characteristics are uniform in two branches of the line of sight, each of  thickness $L/2$ as shown in Figure 5.  The first branch is from the ingress point to an interior point,  and the second branch extends from the interior point to the egress point.  The values of these interior points depend on whether the parameter $\xi < R_1 \equiv R_0 - \Delta$, or  $\xi > R_1$.  In the former case, the interior points are points where the line of sight intersects the inner surface of the shell.  In the latter case, both interior points are the same as  the midpoint of the chord through the shell.  This definition is clarified by reference to Figure 5.  Equation (8) is the fundamental equation in our analysis.

It may easily be shown that the rotation measure through a shell of the sort shown in Figure 5 is of order $RM_0 \equiv 2 C n B_{z0} R_0$.  The maximum value of the rotation measure occurs for chords with $\xi > R_1$, and is $\frac{16}{9}RM_0$ for a compression factor $X=4$.  The difference $\Delta RM$ between the maximum rotation measure and the $RM$ of the central chord depends on the ratio $\frac{R_1}{R_0}$, but is of order $\Delta RM \simeq \frac{4}{3}RM_0$.    
 
\subsubsection{Estimating the Plasma Density in the Shell} Equation (8) takes as its input several properties of the plasma shell illustrated in the cartoon of Figure 5, and perhaps seen in the H$\alpha$ map of Figure 4.  Obviously, one of the most important is the assumed constant plasma density $n$ in the shell.  This parameter can be estimated from the intensity of the H$\alpha$ emission shown in Figure 4.  The relation between the measured H$\alpha$ intensity in Rayleighs, $I_{H \alpha}$ and the emission measure is \citep{Madsen06}
 \begin{equation}
 EM = 2.75 T_4^{0.9} I_{H \alpha}
 \end{equation}
 where $EM$ is in the customary units of cm$^{-6}$-pc, and $T_4$ is the plasma temperature in units of $10^4$ K.  If we adopt a customary value of $T_4 = 0.8$, this relation becomes 
 \begin{equation}
 EM = 2.25 I_{H \alpha}
 \end{equation}
 We measured the $I_{H \alpha}$ values from the WHAM map for each of the lines of sight through the shell to a radio source.  The resultant emission measures showed modest variation from one line of sight to another, with a total range of 192 - 360 cm$^{-6}$-pc.  
 
 It is obvious that interstellar extinction will reduce these emission measures below their true values, with corresponding underestimates of the plasma density in the shell.  We estimated the roll of extinction in the following way.  
 
 \cite{Spangler98} analysed a number of lines of sight through the same part of the Cygnus OB1 association \citep[compare the locations of the sources in Figure 4 of this paper with Figure 6 of ][]{Spangler98}, in that case studying the magnitude of interstellar scattering.  As part of that analysis, values for the emission measure along a number of lines of sight to extragalactic radio sources were determined from radio continuum brightness temperatures, which are not subject to interstellar extinction.  The values for the emission measure along those lines of sight are given in Table 2 of \cite{Spangler98}.  We measured the H$\alpha$ intensity along these same lines of sight, and used them to obtain an estimate of the emission measure from the WHAM data, which was compared with the radio-based emission measures.  For the five sources in Table 2 of \cite{Spangler98}, the ratio $\alpha$ of the radio based $EM$ to the optically-based one ranged from 1.4 to 4.2, with a mean of 2.6. The range in values of $\alpha$ may well represent spatial variations in the extinction from one line of sight to another.  Furthermore, since the beam of the radio telescope used for the radio continuum measurements (Max Planck Institut f\"{u}r Radioastronomie at 
Effelsburg) and WHAM are different, the effective extinction in the two beam solid angles may not be the same. Nonetheless, to obtain a first-order estimate of the effects of extinction in this region, we corrected the WHAM emission measures by the mean value of $\alpha=2.6$. 
 
 Given all of this, our estimate for the density of the shell, calculated along each line of sight, is given by 
 \begin{equation}
 n = \sqrt \frac{2.25 \alpha I_{H \alpha}}{L(\xi)}
 \end{equation}

The density calculated in equation (11) assumes the shell is a spherical annulus with uniform density.  If the plasma is clumped with a filling factor less than unity, the density calculated in (11) will obviously be less than the density in the clumps.  The rotation measure calculated under the assumption of a uniform shell with density given by (11) will exceed the true rotation measure from a shell with a clumped density. It should be noted from equation (11) that the derived density is weakly dependent (proportional to the square root) on the imperfectly-known parameter $\alpha$. 

 For each line of sight, the source-specific value of $\xi$ was used.  Calculation of $L(\xi)$ also requires estimates of the distance to Cygnus OB1, which we take to be 1.8 kpc, and the thickness of the shell.  From the WHAM data shown in Figure 4, we estimated $R_0=71$ pc and $\Delta = 39$ pc. The values for the density which resulted spanned a fairly narrow range of 2.3-3.9 cm$^{-3}$. 
 
 Estimates of the external magnetic field $\vec{B}_0$ were made as follows.  We assumed that the magnitude of the ISM field is $|B_0|=3-5 \mu$G, as indicated by numerous independent studies using Faraday rotation \citep[e.g.][]{Minter96,Haverkorn04,Haverkorn07}. We adopted a value of $B_0=4.0 \mu$G. The orientation of the interstellar magnetic field with respect to the line of sight at the location of the Cygnus OB1 association bubble is unknown, and was left as an adjustable free parameter. 
  
To model the rotation measures for the sample of sources shown in Figure 4, we adopted values of $R_0$, $\Delta$, $n$, and $B_0$ as described above.  Additional free parameters which affect the $RM$s are the angle $\theta$ between the interstellar magnetic field and the line of sight at the location of the Cygnus OB1 association, and an offset constant  $RM_{off}$, which is the background rotation measure in this part of the Galactic plane. Values for the latter two parameters were chosen which gave a good match to the observed rotation measures.  The calculated model $RM$s used the specific value of $n$ for each line of sight as obtained from equation (11).  The model $RM$s thus incorporated available observational information, rather than use a single mean value for $n$ as would be required by strict adherence to the model.  

The parameters of this model for the shell associated with the Cygnus OB1 superbubble are given in Table 2 as ``Model A''. The rotation measures emergent from this model are shown in Figure 6, in the same format as Figure 4.  
 \begin{figure}
 \vspace{0.4cm}
\epsscale{0.60}
\includegraphics[width=25pc]{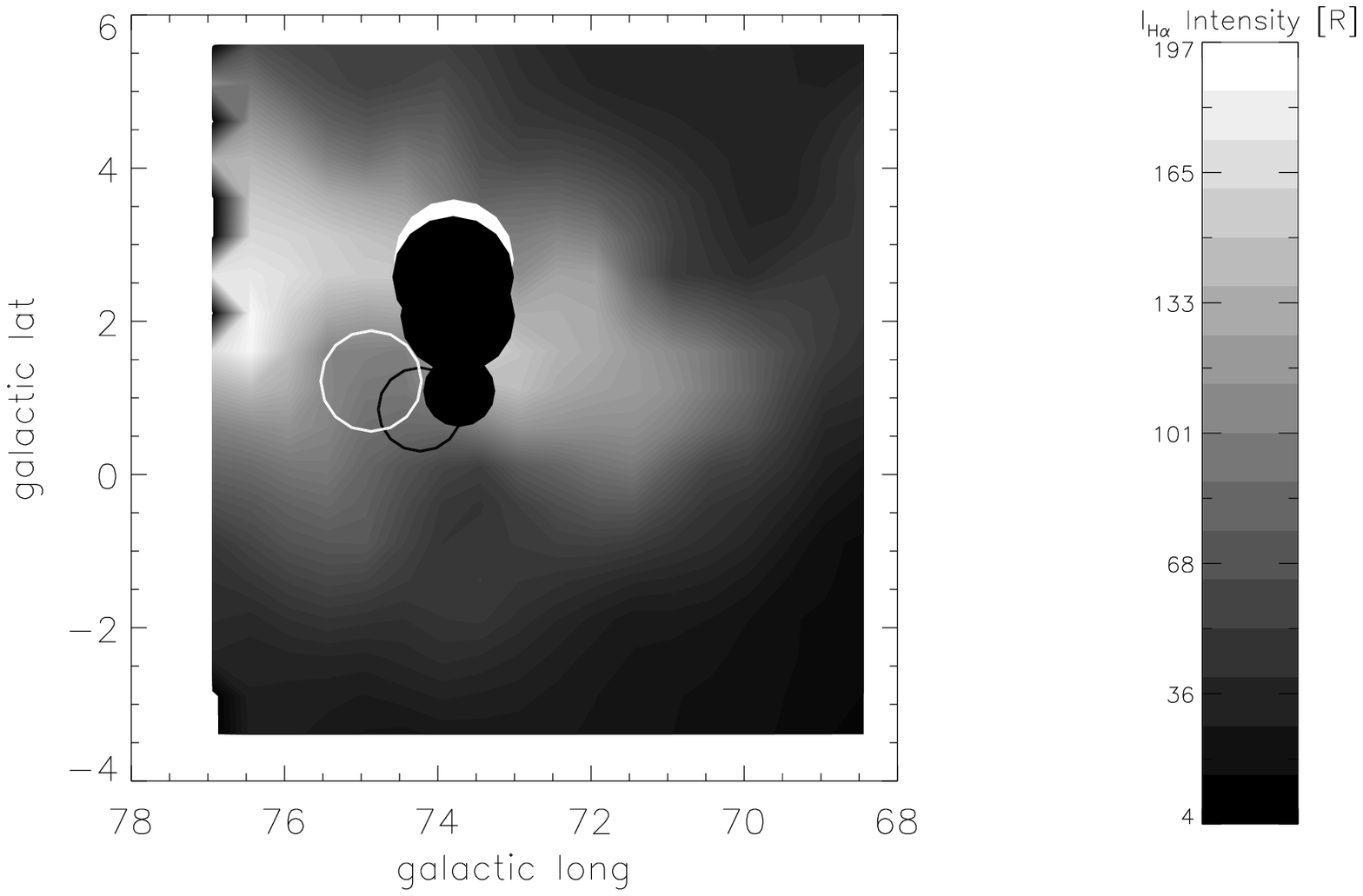}
 \caption{A set of model rotation measures along the observed lines of sight, in the same format as Figure 4.  For this calculation, we used plasma shell densities of 2.3-3.9 cm$^{-3}$, and an external interstellar magnetic field of $4.0 \mu$G oriented at an angle of $59^{\circ}$ with respect to the line of sight. }
 \end{figure}
Figure 6 is qualitatively similar in appearance to the true measurements, given in Figure 4.  The range of model $RM$s is from +665 to -359 rad/m$^{-2}$. We also carried out calculations for a model shell similar to that discussed by \cite{Spangler98}, and based on observations from the literature and theoretical arguments.  This second model (``Model B'') has $R_0=94$ pc and $\Delta=13$ pc. Model B has it own set of associated shell densities obtained from the emission measure measurements, and the free parameters $RM_{off}$ and $\theta$ were adjusted to give a good representation of the observations.  This model also satisfactorily represented the observed range of $RM$s, with the model $RM$ values ranging from +784 to -274 rad/m$^2$.  The parameters for Model B are also given in Table 2. 

The ability of these simplified shell models (with reasonably accurate values of the shell parameters) to reproduce the magnitude and angular scale of the $RM$ values we observe is an important scientific result of this paper. We do not claim that the models presented in Table 2 are unique.  It is obvious from equation (8) that the plasma density, the magnitude of the upstream magnetic field, and the angle between the interstellar field and the line of sight are highly correlated. 

\begin{deluxetable}{ccc}
\tabletypesize{\small}
\tablecaption{Models for Shell Associated with Cygnus OB1 association\label{tbl-2}}
\tablewidth{0pt}
\tablehead{\colhead{Shell Parameter} & \colhead{Model A} & \colhead{Model B}}
\startdata
$R_0$ (parsecs) & 71 & 94\\
$\Delta$ (parsecs) & 39 & 13\\
n (cm$^{-3}$) & 2.3-3.9 & 4.3-5.3\\
$B_0$ ($\mu$G) & 4.0 & 4.0\\
$RM_{off}$ (rad/m$^2$) & -723 & -676\\
$\theta$ (degrees) & 59 & 0\\
\enddata
\end{deluxetable}

The calculations presented above are sufficient to demonstrate that a physical structure similar to the plasma shell associated with the Cygnus OB1 superbubble would produce a rotation measure anomaly similar to that which we observe.  A more realistic model, incorporating additional independent information on the properties of the Cygnus OB1 association superbubble \citep{Saken92,Nichols93,Spangler98} would probably result in better agreement between model observables and observations.  
\subsection{Comparison with Other Anomalies Reported in the Literature}
In this subsection, we discuss the evidence for other, similar features in the sky, which might be physically related to the plasma shell around Cygnus OB1.  Jacques Vallee has for many years reported and discussed $RM$ features on the sky \citep[e.g.][]{Vallee93,Vallee04},  which he refers to as interstellar magnetic bubbles.  In comparison with the object discussed here, the interstellar magnetic bubbles are larger in angular extent, are at higher Galactic latitudes, and induce smaller changes in the background $RM$.  Although \cite{Vallee93,Vallee04} does not model these variations in the same way as we do, he does posit a connection with stellar OB associations. Further consideration of the similarity (and difference) of the objects discussed by Vallee, and the plasma shell associated with the Cygnus OB1 association would be worthwhile.  

An object which is in some ways more similar to that described in this paper is the rotation measure anomaly in the Galactic plane at $l \simeq 92^{\circ}$, discovered by \cite{Clegg92}.  In that paper \cite{Clegg92} suggest that it may be similar to our anomaly in Cygnus.  The $l \simeq 92^{\circ}$ anomaly has been mapped in more detail, and with greater density of sources, by \cite{Brown01}, who identified it as one of three similar anomalous regions (not including Cygnus OB1), although the feature at $l \simeq 92^{\circ}$ is the clearest case in their sample.  The change in $RM$ due to this region is of the order of several hundred to 1000 rad/m$^2$, which is quite comparable to the Cygnus OB1 feature.  The angular extent is also similar, with an angular diameter of $2.5^{\circ}$ \citep{Brown01}.  However, in spite of these similarities between the two features, there are no obvious HII regions or interstellar bubbles similar to Cygnus OB1 along the line of sight to the $l \simeq 92^{\circ}$ anomaly.  Although further attention is clearly necessary, this simple, qualitative astronomical observation would seem to preclude a plasma structure or stellar bubble of the sort we have discussed in Section 4.1.  It may be the case that different types of objects can produce such $RM$ anomalies, a conjecture supported by the catalog of \cite{Vallee04}.  Given the similarity in the Faraday rotation of the Cygnus OB1 and $l \simeq 92^{\circ}$ features, we cannot exclude the possibility that the bubble or plasma shell associated with the Cygnus OB1 association is not responsible for the strong Faraday rotation there.  

The best approach for further illuminating this issue would be to examine OB associations in the direction of the Galactic anticenter, where confusion of multiple objects along the line of sight is not an issue. Such an investigation could include both quantitative measurement of the properties of the H$\alpha$-emitting shells, as well as measurements of, or upper limits to, Faraday rotation anomalies. 
\section{Summary and Conclusions}
The conclusions of this paper are as follows.
\begin{enumerate}
\item Polarization observations at the closely spaced frequencies of 4585 and 4885 MHz with the VLA have confirmed the large, positive Faraday rotation measures for the sources 2007+369, 2009+367, and 2013+361 reported by \cite{Lazio90} on the basis of observations at very different frequencies, and analysed in different ways.
\item The observations in this paper, together with those of \cite{Lazio90} and \cite{Clegg92}, appear to show the presence of a ``Faraday Rotation Anomaly'', in which the $RM$ changes by several hundred radians/m$^2$, or more, over an angular scale of 2-4 degrees.
The $RM$ changes by about 800 radian/m$^2$ between the sources observed in this project, separated by an angular distance of $\sim 2^{\circ}$. Inclusion of other sources observed by \cite{Lazio90} and \cite{Clegg92} leads to differences as large as 1300 radian/m$^2$ over angular separations which are only slightly larger.   
\item This Rotation Measure Anomaly occurs for sources which are viewed through a part of the Galactic plane which contains the Cygnus OB1 association.  H$\alpha$ images from the WHAM telescope show what may be a plasma shell caused by the stellar bubble or superbubble produced by the association.  
\item We have developed a simple physical model of a plasma shell associated with a stellar bubble which appears to be present in Cygnus OB1.  Using independent estimates of the shell size, thickness, and mean density, as well as the strength of the interstellar magnetic field outside the shell,  we have produced $RM$ gradients which are consistent with the observations. 
\item The physical relation of this object to other, somewhat similar or very similar objects reported in the literature is not entirely clear, and would benefit from further research.  
\end{enumerate}
\acknowledgements
This research was supported at the University of Iowa by the National Science Foundation under grant ATM-0354782. Catherine Whiting was supported by the Undergraduate Scholar Assistant program at the University of Iowa. The Wisconsin H-Alpha Mapper is funded by the National Science Foundation through grant AST-0204973 to the University of Wisconsin.

\end{document}